\documentclass[sigconf, nonacm]{acmart}

\newcommand\vldbdoi{XX.XX/XXX.XX}
\newcommand\vldbpages{XXX-XXX}
\newcommand\vldbvolume{14}
\newcommand\vldbissue{1}
\newcommand\vldbyear{2020}
\newcommand\vldbauthors{\authors}
\newcommand\vldbtitle{\shorttitle} 
\newcommand\vldbavailabilityurl{URL_TO_YOUR_ARTIFACTS}
\newcommand\vldbpagestyle{plain} 
\usepackage{enumitem}
\usepackage[x11names]{xcolor}
\usepackage{bm}

\usepackage{xcolor}
\definecolor{skyblue}{RGB}{135,206,235}   
\usepackage{tcolorbox}
\newtcolorbox{mybox}{
    standard jigsaw,
    opacityback=.1,  
    boxrule=1pt,
    left=0pt,
    right=0pt,
    top=0pt,
    bottom=0pt,
    rounded corners,
    arc = 1pt,
    before skip=10pt,
    after skip=5pt,
}
\newcommand{\sys}{Gen-DBA}
\newcommand{\model}{DLA}

\newlist{challenges}{enumerate}{1}
\setlist[challenges]{
  label=\textbf{C\arabic*.},
  leftmargin=*,
}

\begin{document}
\title{
\sys{}: Generative Database Agents [Vision]
}

\author{Yeasir Rayhan and Walid G. Aref}
\affiliation{
  \institution{Purdue University, West Lafayette, IN, USA}
  \city{}
  \country{}
}
\email{{yrayhan, aref}@purdue.edu}

\begin{abstract}
Leveraging Machine Learning to optimize database systems, referred to as Machine Learning for Databases (ML4DB, for short), dates back to the early 1990s, spanning indexing techniques, selectivity estimation, and query optimization. However, the idea has gained mainstream traction following the introduction of learned indexes in 2018, triggering a surge of research spanning learned indexes and cardinality estimators to learned query optimizers, storage layout design, resource management, and database tuning. The current ML4DB optimization landscape is dominated by narrow specialist ML models that are small and are trained on limited training data. Each specialist ML model targets a  single database learning task on a fixed database engine, hardware platform, query workload, and optimization objective. As a result, they fall short in real-world settings, where these factors can vary significantly and evolve over time. This leads to an exponential number of ML models with limited portability and generalization capability, thus limiting the utility of existing ML4DB approaches. We address this limitation with \sys{}, a single general-purpose foundation model for optimizing databases with agentic capabilities. This paper presents the vision for \sys{},  provides a sketch design of how to realize it, and highlights several  research challenges that need to be addressed to fully realize \sys{}.

\end{abstract}

\maketitle

\pagestyle{\vldbpagestyle}
\begingroup\small\noindent\raggedright\textbf{PVLDB Reference Format:}\\
\vldbauthors. \vldbtitle. PVLDB, \vldbvolume(\vldbissue): \vldbpages, \vldbyear.\\
\href{https://doi.org/\vldbdoi}{doi:\vldbdoi}
\endgroup
\begingroup
\renewcommand\thefootnote{}\footnote{\noindent
This work is licensed under the Creative Commons BY-NC-ND 4.0 International License. Visit \url{https://creativecommons.org/licenses/by-nc-nd/4.0/} to view a copy of this license. For any use beyond those covered by this license, obtain permission by emailing \href{mailto:info@vldb.org}{info@vldb.org}. Copyright is held by the owner/author(s). Publication rights licensed to the VLDB Endowment. \\
\raggedright Proceedings of the VLDB Endowment, Vol. \vldbvolume, No. \vldbissue\ %
ISSN 2150-8097. \\
\href{https://doi.org/\vldbdoi}{doi:\vldbdoi} \\
}\addtocounter{footnote}{-1}\endgroup


\section{Introduction}  
\label{sec:introduction}
\underline{Gen}erative \underline{D}ata\underline{b}ase \underline{A}gents, \sys{} for short, 
are envisioned to be 
a single general-purpose database foundation model capable of (i)~performing a broad range of database learning tasks (e.g., query scheduling, query optimization, data layout design, configuration knob tuning), (ii)~controlling diverse database engines (e.g., PostgreSQL, DuckDB, MySQL, Oracle) (iii)~across various  hardware platforms (e.g., Intel, AMD, NVIDIA, Amazon), (iv)~various data and query workloads (e.g., OLTP, OLAP, HTAP, ML workloads), and (v)~supporting multiple optimization objectives (e.g., query throughput, query latency, cost, service level agreements). \sys{} decomposes the database management system (DBMS, for short) into fine-grained components, referred to as Operational Units (OU, for short), and continuously profiles them at run-time with minimal overhead to collect training data. \sys{} formulates each database learning task as a Next Token Prediction problem (NTP, for short)~\cite{BengioDVJ03} 
and employs an Offline Reinforcement Learning (RL, for short) algorithm, namely Decision Transformer (DT, for short), to solve it. At the core of \sys{} is a Database–Language–Action (\model{}) model comprising \textit{three} components: {\em Database}, {\em Language}, and {\em Action}. 
The {\em Database} 
components takes as input the database state as a fixed set of OU profiles. The {\em Language} component takes as input the control instruction and optimization objective in natural language. Finally, the {\em Action} component outputs the database actions by conditioning on the database state, control instruction, and optimization objective. 

\begin{figure}[t]
    \captionsetup{belowskip=-20pt}
    \captionsetup{aboveskip=-4pt}
    \centering
    \includegraphics[width=0.94\columnwidth]{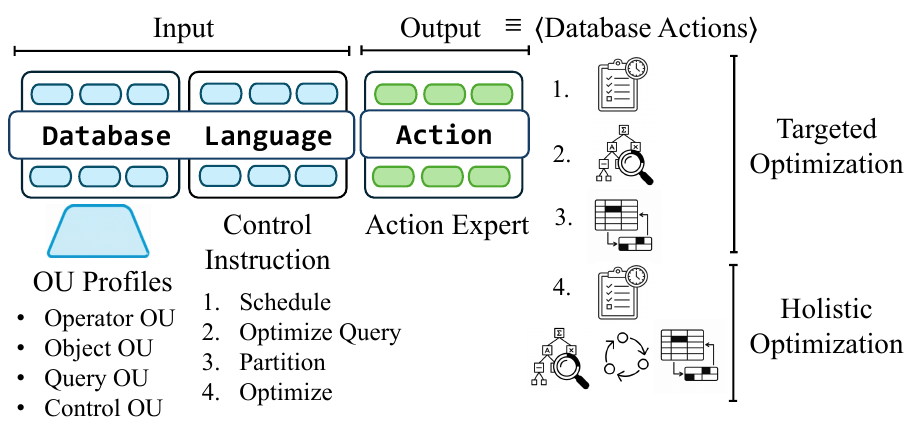}
    \caption{Database Language Action (\model{}) model.}
    \label{fig:intro}
\end{figure}

Refer to Figure~\ref{fig:intro}. \sys{} takes as input a fixed set of OU profiles 
that form the database state. Depending on the control instruction, e.g., \texttt{schedule}, \texttt{optimize query}, or \texttt{partition}, the underlying \model{} model of \sys{} outputs actions for scheduling, query optimization, or partitioning, respectively. We term this \textit{targeted} optimization, where the user explicitly specifies the learning task to be solved. Beyond targeted optimization, \sys{} also supports {\em holistic} optimization, where the user provides a high-level instruction, e.g., \texttt{optimize}, and \sys{} autonomously determines the appropriate actions. These actions may correspond to a single task or a composition of multiple tasks, e.g., scheduling followed by partitioning and then query optimization. 

The envisioned \sys{} can offer several advantages over the current ML4DB approaches for database optimization. \sys{} can replace an exponential number of separate \textit{specialist} ML models with a single unified \textit{generalist} model. \sys{} can eliminate the need to train, deploy, validate and maintain multiple task- and context-specific 
ML models, 
and hence
reduce operational overhead and improve resource efficiency. \sys{} can support portability to unseen configurations. 
\sys{} will maintain a unified action vocabulary for each database learning task and learn from data spanning (1)~different tasks, (2)~different database engines, (3)~different hardware platforms, (4)~various data and query workloads, and (5)~various optimization objectives. These aspects form the \textit{five} axes of database optimization. The underlying ML model of \sys{}, i.e., \model{}, is generative in nature. Thus, \sys{} can discover new policies, contrary to the current ML4DB models that are discriminative. This generative capability, combined with learning across diverse tasks and contexts, allows \sys{} to transfer knowledge both within and across the five axes stated above, improving generalization, and adaptability.

\noindent\textbf{Position.} We argue that {\bf the next wave of ML4DB research should focus on developing a \textit{generalist} database foundation model, in contrast to the narrow specialist ML models that currently dominate the database optimization landscape.} Generalist models are now a reality and given the success of foundation models in the domain of Natural Language Processing (NLP), Computer Vision, and Robotics, we identify \sys{} as a possible next step in realizing a \textit{generalist} approach for database optimization.

\section{Motivating Premise and the Vision}
\noindent \textbf{Database Optimization. }
The database optimization landscape is combinatorial in nature, spanning along five key axes: {\bf (1)~database learning tasks, (2)~database engines, (3)~hardware platforms, (4)~data and query workloads, and (5)~optimization objectives}. For example, looking at 
the top  database systems publication venues, 
since 2020, more than 450 papers have been published targeting more than 27 different \textit{database learning tasks}~\cite{explode-task}. Examples include but are not limited to configuration knob tuning~\cite{LaoWLWZCCTW25,llm-dbtune,LiZLG19,Trummer22,KanellisDKMCV22, ZhangCLWTLC22,AkenPGZ17,ZhangLZLXCXWCLR19, ZhangWCJT0Z021,CeredáVCD21,DuanTB09,KunjirB20, LiZLG19}, physical index design~\cite{AlMamunWHWA26,MarcusKRSMK0K20, FerraginaV20, KipfMRSKK020, KraskaBCDP18, DingMYWDLZCGKLK20, NathanDAK20, DingNAK20,ChatterjeePKI24}, index recommendation~\cite{KossmannKS22, SiddiquiWNC22, PereraORB21, DingDM0CN19, ZhouLZL24}, data layout design~\cite{DingABPJPPPPPSS24,DingMCWLLKGK21,YangCWGLMLKA20,ZapridouMA22,MaddenDKSCMT22}, workload characterization and forecasting~\cite{MaAHMPG18,GaoHZ00C23,GaoHZ00C23,WanZC0LLTSCKK23}, query optimization~\cite{KrishnanYGHS18,MarcusP18,MarcusP19,MarcusNMZAKPT19,YangC0MLS22,ChenGCT23,ChenCLLWZSZ23,ZhuCDCPWZ23,LiuZMS25,Marcus23,ZhangXS25,TaoMJZGM25,YiTIM25,ZhangIM0GLFHPJ22,AnneserTCXPLM23}, query rewrites~\cite{ZhouLCF21,ZhouLCF21,SunZLYFZ25,LiYWCB24}, cardinality and cost estimation~\cite{HeinrichLWKB25,HilprechtB22,KipfKRLBK19,LiangCXYCX024,MarcusP19,SunL19}, etc. 
Similarly, there has been an explosion of \textit{database engines} with distinct architectural choices, and target data models. To date, over 1,057 database engines have been reported~\cite{explode-db}. Furthermore, these database engines are hosted on a wide range of \textit{hardware platforms}. Major cloud providers, e.g., Amazon EC2, Google Cloud, Microsoft Azure provide as many as 1057, 400, 1000 hardware instance choices~\cite{cidr-cloudspecs}, respectively.  

\noindent \textbf{Current ML4DB Premise. }
Most research efforts in the current ML4DB literature design a single ML model to optimize one specific configuration within the five-dimensional optimization landscape~\cite{KrishnanYGHS18,MarcusP18,MarcusP19,MarcusNMZAKPT19,YangC0MLS22,ChenGCT23,ChenCLLWZSZ23,ZhangCPR23,ZhuCDCPWZ23,LiuZMS25,Marcus23,LiuSS25,ZhangXS25,TaoMJZGM25,YiTIM25,ZhangIM0GLFHPJ22,AnneserTCXPLM23,AkioyamenYM24,SabekUK22,MaoSVMA19,LaoWLWZCCTW25,LiZLG19,Trummer22,KanellisDKMCV22, ZhangCLWTLC22,AkenPGZ17,ZhangLZLXCXWCLR19, ZhangWCJT0Z021,CeredáVCD21,HeinrichLWKB25,HilprechtB22,KipfKRLBK19,LiangCXYCX024}. For example, LSched~\cite{SabekUK22} focuses on transaction scheduling in the QuickStep~\cite{quickstep} database engine on Intel Skylake hardware  with the goal of minimizing tail latency for the TPC-H~\cite{tpch} workload. Given the \textit{five} optimization axes, this approach can lead to a combinatorial explosion, resulting in as many as $n^5$ learned model choices. A shift along any axis demands a separate model from scratch. This largely invalidates prior efforts, preventing both within- and cross-axis knowledge transfer across the five axes. This also significantly increases the startup cost of optimizing a database learning task. Not to mention the operational and resource overhead of training, deploying, and maintaining as many as $n^5$ models. 

Additionally, portability remains a key challenge for these narrow-specialist approaches in unseen deployment scenarios, where the hardware platform and data and query workloads may not be known in advance, e.g., as in cloud environments. 
Existing ML4DB models are predominantly discriminative~\cite{Bishop07}, i.e., these models learn the decision boundaries by modeling the posterior $p(y|x)$, where $y$ denotes the label of input $x$. This class of models maps inputs to a predefined label space and cannot synthesize novel solutions. For example, in classification settings, the model is limited to selecting from a fixed set of options, i.e., the available classes. In practice, ML4DB models are  small and are trained on limited data, e.g., MSCN~\cite{KipfKRLBK19} is a 3 MB cardinality estimator. This approach fundamentally contrasts with the scaling-based recipe that has driven the recent breakthroughs in ML, particularly  Large Language Models (LLM, for short).

\noindent \textbf{Vision. }
The current state of ML4DB research closely mirrors that of ML in NLP, Computer Vision, and Robotics roughly five years ago. This is not coincidental, given the relatively late adoption of ML in database systems. In recent years, however, ML has undergone a dramatic paradigm shift, driven by the emergence of Large Language Models (LLM, for short)~\cite{BrownMRSKDNSSAA20,Radford2019LanguageMA,Radford2018ImprovingLU,abs-2302-13971,abs-2303-08774,abs-2401-02954}, Vision Language Models (VLM, for short)~\cite{abs-2407-07726,RadfordKHRGASAM21,AlayracDLMBHLMM22}, and Vision Language Action models (VLA, for short)~\cite{abs-2504-16054,abs-2410-24164,KimPKXB0RFSVKBT24}. These large-scale Transformer-based models, with millions to billions of parameters, are pre-trained on heterogeneous, internet-scale datasets using self-supervised or unsupervised objectives. 
Given the success and widespread adoption of these models across their respective domains, it is evident that ML4DB requires a similar paradigm shift. Rather than continuing along the incremental, task-specific trajectories, where we build \textit{specialist} models pertaining to a single configuration in the five-dimensional optimization landscape, \textit{ML4DB systems should leapfrog directly into the generative, foundational paradigm that can share knowledge across the five axes}. This foundational paradigm enables effective knowledge transfer both within and across axes, reduces startup and operational costs, and improves portability to unseen deployment scenarios. This has the potential to provide a more scalable and sustainable approach to database optimization in general. This vision motivates \sys{}, a step toward realizing a foundation model for optimizing databases.

\noindent\textbf{\sys{}: A Generative Foundation Model with Agentic Capabilities For Databases. }
The underlying model of \sys{}, namely \model{}, is a pre-trained language model (See~\S\ref{sec:dla}) that is \textit{generative}~\cite{NgJ01} in nature, i.e., it learns the joint probability distribution $p(x, y)$ over inputs $x$ and outputs $y$. In practice, this joint distribution is modeled autoregressively via NTP, which factorizes $p(x, y)$ into a sequence of conditional distributions over tokens. Given a tokenized representation of the input, i.e., the database state and previously generated outputs, i.e., the database actions, \sys{} predicts the next output token by conditioning on all prior input and output tokens. This autoregressive formulation allows \sys{} to generate structured database actions rather than selecting from a fixed, predefined label set. As a result, \sys{} can compose novel optimization strategies by combining simple action primitives into variable-length action sequences. Finally, this formulation aligns with the training objective used in LLMs, enabling \sys{} to benefit from the same scaling principles~\cite{scale-ar, scale-neural} while grounding action generation in database semantics.  

While the underlying language model serves as the cognitive core of \sys{}, it is insufficient on its own, as it lacks a motor core that translates high-level decisions into low-level, executable database actions. From a systems perspective, the presence of a motor core is essential, as it enables autonomous operation within the database system. Without a motor core, a foundation model cannot directly observe the database state to drive optimization decisions. Instead, it requires an intermediary, e.g., a human-in-the-loop, to feed database state to the model and model outputs back to the database~\cite{TanZLYPCMZR25,LaoWLWZCCTW25,Trummer22,LaoWLWZCCTW25,HuangLZZYLZCCL25,Zhang25}. This incurs additional engineering overhead, as task-specific action parsers are needed to translate model outputs into executable database actions. Hence, \sys{} integrates a motor core, in the form of a dedicated Action Expert (See~\S\ref{sec:dla}), that enables \sys{} to directly observe, act, and execute the learned database actions within the database system. This eliminates the need for external translation layers, as both the input database state and the output database actions are expressed in the native control interfaces of the underlying ML model and the database engine. 

\noindent\textbf{Scope of \sys{}. }
The concept of foundation models for databases remains broad and open to interpretation. We define a \emph{database foundation model} as one that generalizes across five key axes, i.e., database learning tasks, database engines, hardware platforms, data and query workloads, and optimization objectives. Precisely, \sys{} focuses on database learning tasks that can be expressed as an NTP problem. This includes tasks where optimization decisions can be generated as a sequence of discrete actions, e.g., scheduling, query optimization, configuration knob tuning, resource management, and so on. Tasks that do not naturally fit this formulation, e.g., learned index structures, are outside the scope of \sys{}. Although \sys{} is agentic, its capabilities are intentionally limited to the database system itself. \sys{} can observe database state, predict database actions, and execute them through the database engine's native control interface. It does not perform general agent behaviors, e.g., internet search, external tool use, or interaction with environments outside the database. 

\section{Realizing the vision of \sys{}}
\sys{} draws inspiration from generative foundation models, e.g., LLMs~\cite{BrownMRSKDNSSAA20,Radford2019LanguageMA,Radford2018ImprovingLU,abs-2302-13971,abs-2303-08774,abs-2401-02954}, VLMs~\cite{abs-2407-07726,RadfordKHRGASAM21,AlayracDLMBHLMM22},  VLAs~\cite{abs-2504-16054,abs-2410-24164,KimPKXB0RFSVKBT24}, and Generative Agents~\cite{ParkOCMLB23} to optimize and reason about DBMSs. In this Section, we discuss the key challenges associated with realizing the vision of \sys{} and propose the building blocks that can mitigate them. The challenges are presented in no particular order of importance or difficulty. 

\noindent \textit{Example. } 
Consider the following running example, where the target learning tasks include core scheduling and data partitioning, query optimization, and data layout design. The combination of data partitioning and the core scheduling strategies is referred to as spatial scheduling~\cite{GicevaARH14,pmoss}. The data partitioning strategy decides the location, i.e., the NUMA node of a data partition. The core scheduling policy decides the hardware core, i.e., the core ID responsible for executing an incoming query targeting a data partition. The goal is to learn a scheduling policy for a given database object, e.g., a main-memory B$^+$-Tree. The goal of query optimization is to generate a physical query evaluation plan of an input SQL query. Finally, the goal of data layout design is to find the optimal physical layout of all relational tables in a given schema. The target hardware platforms vary by compute vendors that include Intel, AMD, and NVIDIA. The target database engines include PostgreSQL~\cite{pgsql}, DuckDB~\cite{duckdb}, 
and a research prototype based on a main-memory B$^+$-Tree~\cite{btreecode}. The database schemas and query workloads include YCSB~\cite{CooperSTRS10}, TPC-H~\cite{tpch}, and JOB~\cite{LeisGMBK015}. 
Finally, the optimization objectives include query throughput, average query latency, and tail latency. We use this running example throughout the remainder of this Section to illustrate \sys{'s} building blocks. 

\subsection{Formulating Learning Tasks}
\label{sec:ntp}
\noindent\textbf{Challenges. } Learning task formulation fundamentally shapes the scope and capability of a database foundation model. We present the following key challenges. 
\begin{challenges}[start=1]
  \item How should database learning tasks be formulated to support a foundation model for databases that leverages generative modeling and scalable training on large datasets?
  \item What are the fundamental building blocks of database learning tasks, and how can these building blocks be composed to form expressive and effective optimization strategies that cover the full optimization landscape of a given learning task?
\end{challenges}

\noindent\textbf{Next Token Prediction. }
We envision Next Token Prediction (NTP, for short) to be the primary building block for formulating database learning tasks. NTP is at the core of modern LLMs. By learning to predict the next token in a sequence, LLMs build a deep understanding of the world, and can exhibit emergent behaviors. At a high level, NTP models the probability of the next token conditioned on previously observed tokens. 
This simple yet powerful formulation enables foundation models, e.g., LLMs to scale effectively with increasing model capacity and training data. Motivated by this, we envision \sys{} to cast each database learning task as an NTP problem. This, in turn, requires identifying the fundamental building blocks of each learning task, i.e., its token space. We now return to our running example. For the task of scheduling a main-memory index, the index is logically partitioned into multiple index slices based on the index key. The hardware is modeled as a two-dimensional die, where each cell represents a hardware core with a unique $\langle \texttt{CORE ID}\rangle$. Suppose that there are 6 index partitions and 24 cores. Under the NTP formulation, spatial scheduling reduces to predicting the $\langle\texttt{CORE ID}\rangle$ of the next index partition. Data partitioning is implicit in this formulation, as the $\langle\texttt{NUMA NODE}\rangle$ associated with the selected core determines where the corresponding data is placed. The token space for the scheduling task comprises the available $\langle\texttt{CORE ID}\rangle$s of the target hardware. 

In a similar spirit, for the task of query optimization, the physical query operators of the target database engine define the token space. Under the NTP formulation, this task reduces to predicting the next physical query operator in the query evaluation plan. For example, predicting the next operator sequence, i.e., \texttt{TableScan(C)}, \texttt{TableScan(O)}, \texttt{NLJoin(C, O)} generates the final query plan for the following SQL query: \texttt{SELECT O.* FROM Orders O JOIN Customer C ON C.c\_custkey = O.o\_custkey}. Finally, for the task of data layout design, the $\langle\texttt{COLUMN ID}\rangle$s along with two special tokens, i.e., $\langle\texttt{END GRP}\rangle$ and $\langle\texttt{END PAGE}\rangle$ define the token space. Under the NTP formulation, the data layout design task reduces to predicting the next $\langle\texttt{COLUMN ID}\rangle$ in the database schema. Consider a Table R (a, b, c). Predicting the token sequence $\langle\texttt{Ra,Rb,END GRP,Rc}\rangle$ yields a data morphing layout~\cite{HankinsP03} in which columns \texttt{Ra,Rb} are grouped together and are stored in a single page, while \texttt{Rc} is stored in a separate page. In contrast, predicting the token sequence $\langle\texttt{Ra, END PAGE,Rb,END PAGE,Rc,END PAGE}\rangle$ yields a DSM, i.e., a column-major layout~\cite{dsm}, where each column is stored separately in a separate page. Note that unlike LLMs, where the token space is homogeneous, the token spaces of database learning tasks are inherently heterogeneous. Each database learning task requires its own distinct set of tokens. 



\subsection{Designing the Underlying ML Model}
\label{sec:dla}
\noindent\textbf{Challenges. } The underlying ML model is the core enabler of \sys{}. It dictates how database state is encoded, how optimization decisions are generated, and how the model acts as an autonomous agent. The key challenges are as follows.
\begin{challenges}[resume]
    \item What model architecture is best suited for building a foundation model for databases?
    \item How can a database foundation model be equipped with agentic capabilities to enable end-to-end optimization?
\end{challenges}

\noindent\textbf{Database-Language-Action Model. }
We envision a Database–Language–Action model (\model{}, for short) as the backbone of \sys{}. \model{} integrates database state, control instructions, and database actions into a unified framework. Given the current database state, a control instruction, and an optimization objective specifying the target learning task, \model{} generates executable actions pertaining to the learning task. The design of \model{} is inspired by Vision–Language–Action (VLA) models used in robotics~\cite{Zitkovich23,Brohan23,Kim24,Ghosh24,Liu25,Black24,PhysicalIntelligence25,Ye25}. However, unlike VLAs that process visual observations, e.g., images or videos of a robot’s environment, \model{} operates on database operational unit profiles (See \S\ref{sec:ou}). At the center of \model{} are two Transformer-based components: (1) a pre-trained LLM that 
implements the Database–Language (DL) component of \model{}, and (2) an Action Expert that 
implements the Action component of \model{}. Any pre-trained LLM (e.g., \texttt{Gemma2B}~\cite{GemmaTeam24,GeminiTeam23}) can serve as the cognitive core, determining the overall model capacity and optimization capability of \model{}. To process database state, the Database component places multiple encoders with different architectures on top of the pre-trained LLM. The database state is projected into the same token embedding space as language tokens, leveraging the rich semantic priors embedded in the LLM. The database schema and query workload are processed as language tokens. The Language component provides an interface for expressing both basic and complex control instructions, which are translated into language tokens. The control instructions can specify actions, constraints, and optimization objectives, e.g., ``Optimize query 1a of the JOB~\cite{LeisGMBK015} workload under a 100 MB memory constraint", instructs \sys{} to avoid memory-intensive actions, e.g., hash joins. 

To prevent low-level execution details (i.e., database actions) from polluting high-level decision learning, \model{} employs a separate Transformer (e.g., \texttt{Gemma300M}) as an Action Expert. The Action Expert is trained from scratch and enforces one-way attention isolation, i.e., the action tokens can attend to the full Database–Language context, while the Database–Language backbone is causally masked from attending to action tokens. This ensures a clean separation between high-level reasoning and low-level action generation, where {\em the Action Expert converts cognitive knowledge into executable database actions, enabling \model{} to act as an agent}. Each database learning task in \sys{} maintains a separate action space, defined over the token space introduced in \S\ref{sec:ntp}, and is augmented with a $\langle\texttt{MODE}\rangle$ token to specify the task semantics. Depending on the control instruction (e.g., scheduling), the Action Expert in \model{} is trained to select the appropriate mode, i.e., $\langle\texttt{MODE}=0\rangle$ and predict the corresponding scheduling tokens. 
Note that, \sys{} maintains a unified action vocabulary for each learning task to ensure portability. The action tokens in \sys{} follow a strict action grammar to ensure safety and reliability. For example, in spatial scheduling, the maximum number of action tokens is determined by the maximum number of index partitions and is consistent across indexes and database engines. Together with the OU profiles, this enables \model{} to learn from scheduling policies across multiple database engines and index structures, and to generalize unseen environments. 

\subsection{Abstracting the Database State}
\label{sec:ou}
\noindent\textbf{Challenges. } The abstraction of database state determines the capability, expressive power, and generalizability of \sys{}, which raises the following challenge.
\begin{challenges}[resume]
    \item What database signals are necessary to build a foundation model for databases that can mitigate the impedance mismatch across database engines, hardware platforms, data and query workloads, and optimization objectives?
\end{challenges}

\noindent\textbf{Operational Units. }
We envision the DBMS to be decomposed into fine-grained Operational Units, referred to as OUs~\cite{0006ZJWBLMP21,ChaudhuriN98}. \sys{} categorizes the OUs into four categories, i.e., (1)~Operator OUs: Individual database execution operators, e.g., sequential scan, index scan, index lookup, insert, update, join hash table build, join hash table probe, etc. (2)~Object OUs: Database objects that define data organization and access paths, e.g., database schemas, tables, and indexes. (3)~Query OUs: Individual queries and groups of queries representing the workload. (4)~Control OUs: System-level control parameters and signals. The goal is to cover the essential components of the database stack through these OUs, including the query optimizer, the storage and query execution engine.

For each Operator OU, \sys{} will profile a set of performance events spanning the full hardware stack, i.e., CPU, cache, memory, network, and storage subsystems. Example events include, but are not limited to, CPU cycles, instructions, L1-data cache (L1-D) misses, last-level cache (LLC) misses, memory accesses and misses, local and remote DRAM accesses, per–network-interconnect incoming and outgoing traffic, PCIe read and write bandwidth, disk I/O latency and throughput, etc.  For each Object OU, \sys{} will collect the database schema, index properties, and database statistics. The database schema and index properties are encoded using a domain-specific language (DSL, for short). Table and index statistics are collected from the database catalog, e.g., \texttt{pg\_stats}~\cite{pg-stats} and \texttt{pg\_index}~\cite{pg-index}. For each Query OU, \sys{} will profile the query text (SQL), the logical and physical query plans, and query execution duration, when available. As part of the Control OUs, \sys{} will profile system configuration and performance signals, including database knobs (e.g., working memory, maximum parallel workers), hardware characteristics (e.g., NUMA configuration, CPU frequency), operating system settings (e.g., transparent huge pages, page migration), and end-to-end performance metrics, e.g., throughput, average latency, tail latency, and cost. 


\noindent\textbf{Abstraction across hardware and database engines. }
Different hardware platforms, e.g., AMD, Intel, and NVIDIA expose the same performance events using different names when profiling a given Operator OU.  Therefore, \sys{}  maps hardware-specific counters to a common set of \sys{} features. For example, \sys{} can derive the L1-D cache miss latency feature from \texttt{UNC\_L3\_MISS\_LATENCY}, \texttt{CYCLE\_ACTIVITY.CYCLES\_L3\_MISS}, and \texttt{STALL\_BACKEND\_MEM} on AMD Milan, Intel Skylake, and NVIDIA Grace Hopper servers, respectively. 
Beyond hardware, \sys{} must also generalize the Operator OU profiles across database engines. To this end, \sys{} will normalize each Operator OU profile by input tuple count and size. Hardware events will be further scaled by their data transfer granularity, e.g., cache events are scaled by the cache-line size (64 B), while DTLB and disk events are scaled by the system page size (e.g., 4 KB). The Object OU is encoded using a DSL and a common database catalog. The input SQL is encoded as plain language, and the query execution plans are converted into a \sys{} Intermediate Representation (IR, for short) for abstracting across database engines. Finally, the Control OUs capture specifications of the underlying execution environment. Therefore, they can abstract across different hardware platforms and database engines. Note that, instead of hardware profiles, database metrics (e.g., numbers of reads and writes) could also serve as Operator OU profiles. However, this would require abstraction across thousands of database engines, compared to only a few hardware vendors.

\subsection{Integrating \sys{} Non-intrusively}
\noindent\textbf{Challenges. } It is essential that the underlying \model{} model can be trained continuously and that \sys{} can monitor the database without impacting performance, which raises the following challenges.

\begin{challenges}[resume]
    \item How can a foundational database agent be trained without interfering with database live query execution?
    \item How can a foundational database agent continuously observe and monitor database state 
    with minimum overhead?
\end{challenges}
\noindent\textbf{Offline Reinforcement Learning. }
We envision \sys{} to adopt Offline RL, precisely a Decision Transformer (DT, for short)~\cite{ChenLRLGLASM21} to train the underlying \model{} model. DT isolates the \sys{'s} learning process from the critical path of query execution. \sys{} constrains its RL agent to learn the scheduling policy from a pre-collected dataset (See \S\ref{sec:dataset}), enabling scalable training and improved convergence. This is in contrast to Online RL that continuously interacts with the database via a feedback loop to balance exploration and exploitation that can lead to sub-optimal performance, and is not scalable for large-scale training. DT abstracts sequence modeling as an RL problem, where the sequential decision-making is guided by long-term rewards. It ignores traditional RL techniques, i.e., learning value functions or policies through iterative updates. Rather, it treats RL as a supervised sequence modeling problem, i.e., an NTP problem  that is complementary to the formulation of the learning tasks discussed in \S\ref{sec:ntp}. Precisely, a DT is an auto-regressive sequence model that predicts the next action based on the past observations conditioned on a desired reward. This reward-conditioned nature of DT allows it to align its next-token predictions with optimal decision-making strategies. This aligns well with database learning tasks that typically have a pre-defined objective, e.g., minimizing query latency.  

\noindent\textbf{Low Overhead Profiling. }
We envision the Operator OU profiles in \sys{} as the primary driver of \sys{}’s optimization capabilities, referred to as Profile-Guided Optimization (PGO, for short)~\cite{pgo,awesome_pgo,PettisH90,ChenLM16}. 
The Operator OU profiles provide a unifying interface that links Object, Query, and Control OU profiles with optimization objectives, and enables \sys{} to predict effective optimization strategies. To collect the Operator OU profiles, i.e., the performance events discussed in \S\ref{sec:ou}, with minimum overhead, we envision \sys{} to use the Hardware Performance Monitoring Unit (PMU, for short)~\cite{pmuIntel,pmuAMD,pmuARM}. PMUs are dedicated hardware blocks placed throughout the chip. At the core of these PMU blocks, there are multiple counters (1 to 8 depending on the processor and the location of the PMU block) paired with a control register to track the hardware performance events. During database startup, \sys{} configures the query execution engine to enable profiling, and programs the PMU registers to track the required performance events. During query execution, operators periodically read the pre-configured PMU counters using \texttt{rdpmc} instruction, incurring a low overhead of 9–27 clock cycles~\cite{rdpmc}. Note that PMUs can only capture the CPU, cache, memory, and interconnect behaviors. 
To capture storage behavior, PMU signals can be complemented with OS-level tools, e.g., \texttt{iostat}~\cite{iostats}, to obtain disk-level signals.

\subsection{Training \sys{}}
\label{sec:dataset}
\noindent\textbf{Challenges. } 
One of the key challenges in realizing 
\sys{} is the availability of large-scale training data. While LLMs are pre-trained on massive internet-scale corpora, collecting data of a similar magnitude for database learning tasks remains an open problem. This is unlike NLP, Vision and Robotics, where the availability of large, and open datasets, e.g., 
FineWeb~\cite{PenedoKALMRW024}, Web Language Image~\cite{Chen0CPPSGGMB0P23}, Open X-Embodiment~\cite{ONeillRMGPLPGMJ24}, has been a key enabler of foundation models. This raises the following challenges. 
\begin{challenges}[resume]
    \item How can we construct a large-scale and diverse 
    training dataset for optimizing database systems? 
    \item What training recipe is best suited for building a foundation model for databases?
\end{challenges}

\noindent\textbf{Generating Training Dataset. }
We envision \sys{} to generate training data, i.e., the OU Profiles (See \S\ref{sec:ou}) during normal query execution in the database engine. To construct the training dataset, \sys{} first selects configurations across the five axes. Consider the running example. Assume that \sys{} selects spatial scheduling as the target learning task, a B$^+$-Tree as the target engine, Intel as the hardware platform, YCSB as the target workload, and the throughput as the optimization objective. The YCSB workload is executed on the B$^+$-Tree, during which worker cores profile Index Operator OUs with minimal overhead (See \S\ref{sec:ou}). These OU profiles are periodically collected by sweeper cores and are offloaded to form the training dataset. The data and query workloads would guide the training data generation process of \sys{}. Representative examples include TPC-H~\cite{tpch} and TPC-DS~\cite{tpcds}, JOB~\cite{LeisGMBK015}, JOB-Complex~\cite{abs-2507-07471}, JOB-Extended, CardBench~\cite{abs-2408-16170} and SemBench~\cite{sembench}. The LLMs can serve as a powerful source for generating query workloads, e.g., SQLStorm~\cite{SchmidtLBN25}. These benchmarks provide diverse workloads that are instrumental for generating training data for \sys{}. Besides, the pre-trained LLM of the underlying \model{} model reduces the requirement of internet-scale training data as the LLM is already pre-trained on large amounts of database-related semantics expressed in natural language.

\noindent\textbf{Representing Training Dataset. }
The training dataset will be ordered into sequences of states $(s_t)$, actions $(a_t)$, and rewards $(r_t)$. Each training sample corresponds to a trajectory $\bm{\tau}$, represented as an interleaved sequence of three types of token, reward-to-go (RTG, for short), state, and action tokens. $\bm{\tau} =\langle \hat{r}_0, s_0, a_0, \hat{r}_1, s_1, a_1, \ldots, \hat{r}_{|\bm{\tau}|}, s_{|\bm{\tau}|}, a_{|\bm{\tau}|}\rangle$. An RTG token $\hat{r}_t$ denotes the future cumulative reward expected from a given timestep $t$ onward. The initial RTG $\hat{r}_0$ is equal to the reward of the sequence. In \sys{}, the database state and the control instruction, the optimization signal and the corresponding action for a database learning task would form the state, RTG, action tokens, respectively. 

Consider the running example of scheduling a main-memory B$^+$-Tree partitioned into six partitions. Assume that the current scheduling policy is $\langle\texttt{C1}, \texttt{C2}, \texttt{C3}, \texttt{C4}, \texttt{C5}, \texttt{C6}\rangle$, where $\texttt{C}_n \in [1, 24]$ denotes the $\langle\texttt{CORE ID}\rangle$ assigned to the $n$-th partition. 
The resulting training sample is represented as follows. At Time $t_0$, $\hat{r}_0=100$ denotes the desired query throughput of the current scheduling policy. $s_0 = \langle \texttt{OP}_{1,\ldots,n_1}, \texttt{OB}_{1,\ldots,n_2}, \texttt{Q}_{1,\ldots,n_3}, \texttt{CO}_{1,\ldots,n_4}, \texttt{IN}_{1,\ldots,n_5} \rangle$ denotes the database state at Time $t_0$, where \texttt{OP}, \texttt{OB}, \texttt{Q}, \texttt{CO}, and \texttt{IN} refer to the Operator, Object, Query, Control OU profiles, and the control instruction, respectively. The Action Token $a_0 = \texttt{C0}$ schedules the first index partition. At Time $t_1$, the RTG and the state tokens are updated to $\hat{r}_1=80$ and $s_1 = \langle \texttt{OP}_{1,\ldots,n_1}, \texttt{OB}_{1,\ldots,n_2}, \texttt{Q}_{1,\ldots,n_3}, \texttt{CO}_{1,\ldots,n_4}, \texttt{IN}_{1,\ldots,n_5}\rangle$, respectively, reflecting the scheduling of the first index partition. Action token $a_1 = \texttt{C1}$ schedules the second index partition. For a given learning task, only the corresponding Operator OU profiles would be activated. The remaining Operator OU profiles would be masked to isolate the dynamics of the target task. For example, during spatial scheduling, query optimization, and the partitioning tasks, $\langle\texttt{OP}_{\texttt{IDX}}\rangle$, $\langle\texttt{OP}_{\texttt{QUERY}}, \texttt{OB}_{\texttt{QUERY}}\rangle$, $\langle\texttt{OP}_{\texttt{TBL\_SCAN}}\rangle$ would be profiled, respectively along with the Query and Control OU profiles, $\texttt{Q}$, $\texttt{CO}$ and control instruction, $\texttt{IN}$, when available. 


\noindent\textbf{Three-stage Training Pipeline. } 
We envision the training pipeline of \sys{} to comprise \textit{three} stages. (1)~Pre-Training: In this stage, \sys{} is trained on large-scale datasets spanning all five optimization axes. The training data include both high-quality and suboptimal examples. This enables the model to learn to distinguish effective strategies from ineffective ones. (2)~Mid-Training: During mid-training, \sys{} is trained on only high-quality data across all five axes. This stage also introduces new system capabilities, e.g., previously unseen query operators. (3)~Post-Training: During post-training, \sys{} is 
to be 
trained on deployment-specific data to align the model to the target execution environment. For example, when deployed on PostgreSQL and Intel, \sys{} will be fine-tuned using PostgreSQL- and Intel-native training data.

\subsection{Optimizing Databases Holistically}
\noindent\textbf{Challenges. } A foundation model for databases should have the capability of performing holistic optimization by coordinating across multiple learning tasks, which raises the following challenge.
\begin{challenges}[resume]
    \item How can a database foundation model move beyond targeted optimizations to perform holistic optimization? 
\end{challenges}
\noindent\textbf{Holistic Optimization. }
We envision \sys{} to accept a general control instruction, e.g., \texttt{Optimize}, and predict actions across multiple learning tasks. Given the instruction \texttt{Optimize} for the JOB workload, \sys{} can recommend reorganizing the physical layout of table \texttt{title}, reschedule the index on \texttt{title.id}, and reoptimize the query evaluation plan for Query \texttt{1a}. To enable this, \sys{} must reason jointly across multiple learning tasks and prioritize actions that maximize the overall reward. The key step will be to augment the existing per-task training dataset with samples that capture the cumulative reward of executing a sequence of database actions across multiple learning tasks. This requires identifying an optimal mixture of training samples that improves \sys{'s} holistic optimization performance while preserving \sys{'s} performance on individual tasks. Recall that the database state, i.e., $s = \langle \texttt{OP}_{1,\ldots,n_1}, \texttt{OB}_{1,\ldots,n_2}, \texttt{Q}_{1,\ldots,n_3}, \texttt{CO}_{1,\ldots,n_4}, \texttt{IN}_{1,\ldots,n_5}\rangle$, 
is shared across all learning tasks. For individual tasks, irrelevant OU profiles are masked (zeroed out). In contrast, for holistic optimization, the full end-to-end database state across all OU profiles is exposed. Similarly, the reward is defined by end-to-end database engine performance rather than by task-oriented rewards. Action tokens represent a sequence of actions spanning multiple learning tasks. For the above example, $\mathcal{A}=\langle\texttt{MODE=2}\rangle\langle \texttt{P}_{1, \ldots,}\rangle\langle\texttt{MODE=0}\rangle\langle \texttt{C}_{1, \ldots,}\rangle\langle\texttt{MODE=1}\rangle\langle \texttt{Op}_{1, \ldots,}\rangle$ form the action tokens, where \texttt{P}, \texttt{C}, \texttt{Op} denote the individual partitioning, scheduling, and query optimization action tokens. 




\section{Related Work}
Wehrstein et al.~\cite{foundational-cidr} propose a vision for database foundation models consisting of multiple pre-trained experts that are trained separately. This design can cause error propagation across experts, leading to performance degradation. Multi-Task Meta-Learning Framework~\cite{WuYYZHLLZZ22} proposes a shared embedding space for different database learning tasks and creates multiple models from the shared space for each corresponding task. Instance-optimized database systems~\cite{KraskaABCKLMMN19,DingMKNNVRK22,Kraska21, Kraska00MNWY23, YuWKLMNMK24, HilprechtB22, ParkTCM17} focus on developing separate ML models for each database component. Self-designing data systems~\cite{Idreos25,IdreosZADHKGMQW18,IdreosZHKG18,IdreosDQAHRLJGL19} optimize its physical data structures according to changes in the hardware, data and query workload. Self-driving data systems~\cite{PavloAALLMMMPQS17,andy-blog,Pavlo21,llm-dbtune,ZhangLBP24} aim to automatically tune database configuration knobs and optimize system behavior. Zero-shot data systems~\cite{HilprechtB22,AgnihotriKBL23, HilprechtB22, HeinrichLKB22, AgnihotriKSHBL24} propose a single model that can generalize across diverse data and query workloads for a given learning task. {\bf All these works only address a subset of the five axes of the optimization landscape and overlook the crucial aspect of leveraging the semantic knowledge from pre-trained LLMs.} 
In contrast, 
\sys{} 
proposes a single unified model that is trained end to end to improve database performance. \sys{} is generative, which makes it capable of synthesizing new strategies.

\section{Conclusion}
\label{sec:conclusion}
In this paper, we present the core vision for \sys{}, a database foundation model with agentic, i.e., autonomous decision-making capabilities. We introduce several challenges that pose key hurdles and outline the corresponding building blocks, namely: Next Token Prediction, Database-Language-Action model, Operational Unit, Hardware PMU, Offline Reinforcement Learning, a three-stage training pipeline, and holistic optimization. Together, these concepts lay the foundation for a new generation of ML4DB systems that move beyond specialist models toward generalist systems.

\balance{}
\bibliographystyle{ACM-Reference-Format}
\bibliography{sample}

\end{document}